# Electron counting in solids: oxidation states, partial charges and ionicity

Aron Walsh[†,*], Alexey A. Sokol[‡], John Buckeridge[‡], David O. Scanlon[‡], C. Richard A. Catlow[‡]

[†]Department of Materials, Imperial College London, London SW7 2AZ, UK
[‡]Department of Chemistry, University College London, London WC1H 0AJ, UK
 E-mail: a.walsh@imperial.ac.uk

The oxidation state of an element is a practically useful concept in chemistry. IUPAC defines it as "*the charge an atom might be imagined to have when electrons are counted according to an agreed-upon set of rules*"[1]. Once the composition of a compound is known, a trained chemist will immediately infer the oxidation states of its components, and in turn anticipate the structural, electronic, optical and magnetic properties of the material. This is a powerful heuristic tool.

In the modern era of quantum chemistry, can the use of formal oxidation states be justified? Let us take the example of $TiO_2$, a popular metal oxide. Following the standard rules, an oxidation state of Ti(IV) is assigned. The formal electronic configuration of the ions are O $2p^6$ and Ti $3d^0$, which is reflected in the band structures calculated from first-principles quantum chemical approaches, i.e. a closed-shell electronic structure where the valence band is formed predominately of filled O 2p orbitals and the conduction band is formed of empty Ti $3d$ orbitals[2]. A detailed theoretical analysis of chemical bonding in rutile $TiO_2$, bridging ionic and molecular orbital models, was provided by Burdett[3]. In further support of the oxidation state assignments, if electrons are added to the material (chemical reduction) they localise on Ti to form a Ti(III) $d^1$ centre, while if electrons are removed (chemical oxidation) they form holes on O ions[4]. Indeed, oxidation via formation of oxygen interstitials in $TiO_2$ yields a peroxy-species in the lattice.[4] Alternatively, if we grow oxygen sub-stoichiometric $TiO_2$, Ti(III) species are observed, and on increase of sub-stoichiometry a $Ti_2O_3$ phase separation is seen. It may therefore be surprising that a charge state of Ti 2.5+ in $TiO_2$ can be assigned on the basis of recent density functional theory (DFT) calculations[5].

The electron density in solids is routinely calculated using a range of techniques in computational chemistry and measured in diffraction experiments. However, the charge on a given ion is not an observable, but relies on a *choice* of model or theory to partition the electron density between atomic centres; and there is in general no unambiguous way to do such partitioning. The wave functions of electrons in crystals are multi-centred and delocalised (Bloch waves) as required by quantum mechanics. Charges could be assigned unambiguously for the special case that electron density does not overlap, which is, however, uncommon; a point explored in detail within the review by Catlow and Stoneham[6]. Through topological analysis of the electron density, and by separation into atom-centred basins using the scheme proposed by Bader[7] (one of many projection schemes[8]), a partial charge of Ti 2.5+ is assigned in $TiO_2$. One problem with Bader's approach is that atoms or ions do overlap. Partial charges such as these are peculiar quantities in chemistry as oxidation and reduction processes involve changes in integral numbers of electrons. Polarisation of an anion towards a cation is impossible to distinguish from a charge transfer on forming a heteropolar bond between such two ions. As such, partial charges can be misused to infer the ionic character of a chemical bond[9]. We note, however, that relative changes in partial charges (in similar chemical environments) have utility for probing chemical processes, e.g. surface catalytic reactions.

What can be measured? In solid-state thermochemistry, charges corresponding to formal oxidation states are consistent with lattice energies from thermochemical data (Born-Haber cycles) for a plethora of inorganic compounds[10]. The response to electromagnetic fields is determined by the dielectric screening, which can be accurately described by models of polarisable ions with formal charges, as can the interatomic forces and hence the phonon dispersion[6]; and such an assignment has equal validity with those based on calculated or measured charge densities. The unambiguous assignment of oxidation states can be made from experiments, e.g. core-level photoemission spectroscopy, based on reference to the corresponding ions in (usually aqueous) solutions. Moreover, electrochemical experiments allow one to count charges going to electrodes directly, proving the (complex) reality of formal oxidation states, while optical techniques can excite electron-hole pairs with particular degrees of localisation. Electron spin resonance (ESR) and nuclear magnetic resonance (NMR) can then be employed to detect localised states on defects that can often be ascribed to particular ions with demonstrably one-(full)-electron character, e.g. reduced Ti(III) centres in $TiO_2$[11].

The classification of chemical bonding in solids as covalent (e.g. Si or diamond) or ionic (e.g. LiF or MgO) is one that continues to promote debate. For any compound containing two or more elements with differing electronegativity, the bonding is heteropolar, which can be described using the language of covalency (i.e. hybridisation of orbitals) or ionicity (i.e. polarisation of ions). These are two alternate descriptions of the same reality[7]. From an ionic perspective, moving from a $TiO_2$ molecule to the solid we do not change the oxidation state, but rather enhance the polarity of bonds. As the coordination of oxygen increases, the Madelung potential stabilises the oxide anion, increasing the ionisation potential of the crystal[12]. Due to this bond polarity, the surface stability of metal oxides is dictated by classical electrostatics[13], a key example is the (Tasker type II) (110) termination of rutile $TiO_2$, which is the dominant crystal facet that does not require complex reconstructions or chemical passivation.

Electron counting in solids is more challenging than it would first appear. The contribution of a particular atom to the electronic structure of a compound is masked by wave function-based quantum mechanics. However, the theory of electron separability and electron groups does provide a solid basis for understanding the oxidation state of an atom in a compound[14] – it tells us how to count electrons, or rather that when we count electrons using standard chemical rules (cf. the VSEPR model) about the valence, electron pairs, etc., we do the right thing. It is our view that absolute values of partial charges should be interpreted and used with caution; the charge assigned can never be definitive and depends on the type of property studied and the type of analysis performed. There has been recent progress in the area, such as recovering integral oxidation states from first-principles within the modern theory of polarisation[15]. Careful analysis can be used to avoid unphysical conclusions such as the Ti(III) nature of Ti in stoichiometric $TiO_2$.

## Acknowledgements

We acknowledge support from the Royal Society, Leverhulme Trust, and the EPSRC (Grant no. EP/K016288/1, EP/J017361/1 and EP/N01572X/1).